\documentclass[a4paper,11pt]{article}
\pdfoutput=1 

\usepackage{jheppub} 

\usepackage[T1]{fontenc} 
\usepackage{amsmath}
\usepackage{amssymb}
\usepackage{amsfonts}
\usepackage{graphicx}
\usepackage{enumitem}
\usepackage{bm}
\usepackage{rotating}
\usepackage{hyperref}
\usepackage{pbox}
\usepackage{xcolor}
\usepackage{array}
\usepackage{physics}
\usepackage{mathtools}
\usepackage{leftidx}
\usepackage{lmodern}
\usepackage[normalem]{ulem}
\usepackage{braket}
\usepackage{tensor}
\usepackage{mathrsfs}
\usepackage{float}
\usepackage[title]{appendix}

\usepackage{tikz}
\usetikzlibrary{arrows.meta,decorations.pathmorphing,math}

\definecolor{plum}{rgb}{0.36078, 0.20784, 0.4}
\definecolor{chameleon}{rgb}{0.30588, 0.60392, 0.023529}
\definecolor{cornflower}{rgb}{0.12549, 0.29020, 0.52941}
\definecolor{scarlet}{rgb}{0.937, 0.161, 0.161}
\definecolor{brick}{rgb}{0.64314, 0, 0}
\definecolor{sunrise}{rgb}{0.80784, 0.36078, 0}

\title{\boldmath Entanglement Harvesting with Moving Mirrors}

\author[a,b]{Wan Cong,}
\author[a,b]{Erickson Tjoa,}
\author[a,b]{and Robert B. Mann}

\affiliation[a]{Department of Physics and Astronomy, University of Waterloo,\\ Waterloo, Ontario, N2L 3G1, Canada}
\affiliation[b]{Institute for Quantum Computing, University of Waterloo,\\ Waterloo, Ontario, N2L 3G1, Canada}

\emailAdd{wcong@uwaterloo.ca}
\emailAdd{e2tjoa@uwaterloo.ca}
\emailAdd{rbmann@uwaterloo.ca}

\abstract{We study the phenomenon of entanglement extraction from the vacuum of a massless scalar field in $(1+1)$ dimensional spacetime in presence of a moving Dirichlet boundary condition, i.e. \textit{mirror spacetime}, using two inertial Unruh-DeWitt detectors.  We consider a variety of non-trivial trajectories for these accelerating mirrors and find (1) an entanglement inhibition phenomenon similar to that recently seen for black holes, as well as (2) trajectory-independent entanglement enhancement in some regimes. We show that the qualitative result obtained is the same for both linear and derivative couplings of the detector with the field.}

\begin{document} 
\maketitle
\flushbottom

\section{Introduction}
    The study of quantum entanglement has far-reaching consequences in many fields, such as in the study of black hole entropy \cite{Sorkin1986entropy,Srednicki1993area} and the    
    anti-de Sitter/conformal field theory (AdS/CFT) correspondence \cite{ryutaka2006holography}. In  more formal algebraic quantum field theory, it was shown that the vacuum state of a quantum field can maximally violate   Bell's inequalities \cite{summers1985bell}. 
    A more operational approach to study   quantum field entanglement was initiated in \cite{Valentini1991nonlocalcorr}, where it was shown that atoms initialized as uncorrelated states can become entangled after some time due to the global nature of field correlators. This particle detector model, now known as the Unruh-DeWitt (UDW) model \cite{PhysRevD.14.870,DeWitt:1980hx}, has been extensively used to study the phenomenon of entanglement harvesting \cite{Salton:2014jaa,Pozas2015}, discerning the topology of spacetime \cite{Smith2016topology} and distinguishing a thermal bath from an expanding universe at the same temperature \cite{Steeg2009}, among others. This operational approach has the advantage that it is easy to apply even in curved spacetimes.
    
    Here we study the phenomenon of entanglement harvesting in the context of \textit{mirror spacetimes} i.e. the Minkowski spacetime spatially divided by a (possibly dynamical) Dirichlet boundary condition, in $(1+1)$ dimensions. This would constitute the first investigation of entanglement harvesting in a highly  non-stationary background quantum field essentially due to the Dynamical Casimir effect (DCE). The DCE has been recently observed experimentally \cite{Delsing2011dynCasimir, Lahteenmaki4234} thus motivating the use of mirror spacetimes to study aspects of particle creation in quantum field theory. 
    
    The conformal invariance of the massless wave equation in $(1+1)$ dimensions, together with the ease of obtaining exact analytic expressions for the Wightman functions of moving mirrors in certain classes of trajectories, make the study of mirror spacetimes a very attractive toy model to gain insights into the physics of quantum fields in curved spacetimes. The study of responses and transition rates of UdW detectors in receding mirror spacetimes was first investigated decades ago \cite{birrell1984quantum, suzuki:1997boundary, davies:1989boundary}. For some classes of trajectories, one can even match certain mirror trajectories to Hawking radiation associated with black hole spacetimes, whether at the level of the response rate \cite{Hodgkinson:2013tsa} or at the level of the Bogoliubov transformation and stress-energy tensor \cite{CarlitzWilley1987reflection}. 
    More recently it has been shown that for some generic trajectories certain limits can also be taken to obtain thermal responses \cite{Good2013mirror} or even model black hole collapse from a null shell \cite{Good2016blackholemirror3,Good2017,Good2018Evanescent}.  We note in this context that a study of entanglement harvesting in black hole spacetimes was recently initiated for the first time for $(2+1)$-dimensional black holes \cite{Henderson:2017yuv},
    and so it is of further interest to investigate entanglement harvesting in spacetimes with non-trivial boundary conditions
    that could simulate the collapse of matter into a black hole.

    We will see that, despite the differences between the physics of mirrors and black holes, a pair of atoms can experience  (1) entanglement inhibition and even \textit{entanglement death} near a mirror, an effect also associated with the event horizon of a black hole  \cite{Henderson:2017yuv}, as well as (2) trajectory-independent entanglement enhancement in some regimes. 
    
    Our paper is organized as follows. In section~\ref{sec: setup} we describe the setup involving the UDW model and provide several \textit{ray-tracing functions} for the mirror trajectories that will be studied in this paper. In section~\ref{sec: results} we study entanglement harvesting between two detectors in the presence of a static mirror and also a non-inertial mirror first studied in \cite{CarlitzWilley1987reflection}. In section~\ref{section: blackhole} we compare these results with that of a mirror trajectory recently described in \cite{Good2016blackholemirror1} that mimics the null black hole collapse model. In section~\ref{sec: distancedep} we show that the same qualitative results are obtained using a derivative-type coupling, before finally concluding in section~\ref{sec: conclusion}. 
    
    In this paper we adopt the natural units $c=\hbar=1$.

    \section{Setup}
    \label{sec: setup}
    In this section we recall the standard UDW model in the context of a massless scalar field in (1+1) dimensions interacting with a first quantized atom. In the Dirac picture, the light-matter interaction is provided by the interacting Hamiltonian of the form
    \begin{align}
    \label{eq: int}
        H_I^j(\tau) = \lambda\chi_j(\tau)\hat\mu_j(\tau)\otimes\hat\phi(\sx_j(\tau))\,, \hspace{0.5cm} j = A,B
    \end{align}
    where $\lambda$ is the coupling strength\footnote{Note that in $(1+1)$ dimensions, the coupling constant is not dimensionless but has units of inverse length in natural units. Here we say that a small coupling constant means the dimensionless  quantity $\lambda\sigma$ is small, where $\sigma$ is the width of the Gaussian switching function.}, $\chi(\tau)$ is a switching function,  $\hat{\phi}$ is the field operator and $\sx_j(\tau)=(t(\tau),x_j(\tau))$ is the trajectory of the atomic detector. In addition, $\hat\mu_j(\tau)=\hat\sigma_j^+e^{i\Omega_j \tau}+\hat\sigma_j^- e^{-i\Omega_j \tau}$ is the monopole moment of the detector, where $\sigma^\pm$ are the ladder operators of the $\mathfrak{su}(2)$ algebra and $\Omega_j$ is the energy gap of the two-level atom. The index $j=A,B$ indicates which atom (UDW detector) we are considering: in the case of entanglement harvesting, we have two atomic detectors so the full interacting Hamiltonian takes the form
    \begin{equation}
        H_I = H_I^A + H_I^B\,.
    \end{equation}
    We shall also consider for simplicity the case where the two atoms are identical so $\Omega=\Omega_A=\Omega_B$. In order to not cleanly account for relative motion between mirrors and the detectors, we focus on the case when both detectors are static in the quantization frame $(t,x)$, so that the proper time for both detectors at rest relative to each other is given by $\tau=t$.
    
    We consider the initial state of the full system to be a separable state $\rho_0 = \ket{\psi}\bra{\psi}$ where $\ket\psi = \ket{0}\ket{g}_A\ket{g}_B$, and where $\ket{0}$ is the field vacuum and $\ket{g}$ is the atomic ground state. The final state of the detector $\rho_{AB}$ is obtained by tracing out the field state after time evolution:
    \begin{equation}
    \begin{split}
        \ket{\psi_f} &= \mathcal{T}\exp\rr{-i\int \dd t H_I(t)}\ket{\psi} := \hat U\ket\psi\,,\\
        \rho_{AB} &= \Tr_\phi \ket{\psi_f}\bra{\psi_f}\,.
    \end{split}
    \end{equation}
    
    In the $\{\ket{g},\ket{e}\}$ basis, the joint state of the detectors will be given by the matrix \cite{Smith2016topology}
    
    \begin{equation}
        \rho_{AB} = \begin{pmatrix}
        1-P_A-P_B & 0 & 0 & X^*\\
        0 & P_B & C & 0 \\
        0 & C^* & P_A & 0\\
        X & 0 & 0 & 0
        \end{pmatrix}+O(\lambda^4)\,.
    \end{equation}
    The expressions for $X,P_j$ (we are not using $C$ in this paper) are given by
    \begin{align}
        X &= -\lambda^2\iint \dd t\,\dd t' \chi_A(t)\chi_B(t') e^{-i\Omega(t+t')}\notag\\
        &\hspace{1.5cm}\bigg[\Theta(t'-t)W(\sx_A(t),\sx_B(t'))+\Theta(t-t')W(\sx_B(t'),\sx_A(t))\bigg]\,,\label{eq: nonlocal}\\
        P_j &= \lambda^2\iint  \dd t\,\dd t' \chi_j(t)\chi_j(t') e^{-i\Omega(t-t')}W(\sx_j(t),\sx_j(t'))\,\label{eq: probability} 
    \end{align}
    where $W(\sx,\sx')=\braket{0|\hat\phi(\sx)\hat\phi(\sx')|0}$ is the pullback of the Wightman function to the detector trajectories and $\Theta(\cdot)$ is the Heaviside step function. Since the detectors are identical, we also choose the same Gaussian switching function $\chi(t)$ for both detectors,
    \begin{equation}
    \label{eq: gaussswitch}
        \chi_j(t) = \exp\left[ -\frac{(t-t_j)^2}{2\sigma^2}\right]
    \end{equation}
    where the parameter $\sigma$ characterises the length of the switching function and hence the duration of interaction with the field and $t_j$ is the temporal peak of the switching function.
    
    The entanglement measure we use here is concurrence $\mathcal{C}(\rho_{AB})$, computed in this case to be \cite{Smith2016topology} 
    \begin{equation}
        \mathcal{C}(\rho_{AB}) =2\max\left\{0,|X|-\sqrt{P_AP_B}\right\}+O(\lambda^4) 
    \end{equation}
    which is simple and transparent for our purposes\footnote{Negativity $\mathcal{N}$ is another usable well-known entanglement measure but in this context does not yield qualitatively different results; furthermore concurrence  admits a simpler interpretation due to a  clean separation of local and nonlocal terms and the fact that it monotonically increases with the entanglement of formation.    }.
    
    In (1+1) dimensions, the metric for Minkowski spacetime can be written in terms of double null coordinates $u,v=t\mp x$ so that
    \begin{equation}
        \dd s^2 = \dd t^2-\dd x^2 = \dd u\,\dd v\,.
    \end{equation}
    The Wightman function for massless scalar field in $(1+1)$ dimensions in terms of these coordinates is 
    \begin{equation}
    \label{eq: freeW}
        W_f(\sx,\sx') = -\frac{1}{4\pi}\log\Bigg[\Lambda^2(\epsilon+i\Delta u)(\epsilon+i\Delta v)\Bigg]
    \end{equation}
    where the logarithm is taken with respect to the principal branch, the subscript $f$ denotes `free space', $\Lambda$ is an IR cutoff and $\epsilon$ is a small positive constant serving as the UV cutoff. In free space, the IR regulator cannot be removed, which is a peculiarity of $(1+1)$ dimensions alone \cite{Eduardo2015firewall}. This leads to the well-known IR ambiguity in the response of a detector coupled linearly to $(1+1)$ massless scalar field.
    
    Unlike the free-space case, the presence of a (moving) mirror i.e. Dirichlet boundary condition removes this IR ambiguity via the ``method of images''.  Hence our investigation  does not suffer the fundamental IR cutoff problem encountered in the free-space scenario.  It modifies the Wightman function via \textit{ray-tracing functions} $p(u)$ or $f(v)$ depending on whether we use $u$ or $v$ to `trace' rays: in the presence of a mirror, some of the `reflected' right-moving modes can be written in terms of the incoming left-moving modes. Anticipating our results, we choose $p(u)$ to avoid issues involving coordinate singularities (see e.g. \cite{CarlitzWilley1987reflection} or \cite{Good2013mirror}) and the Wightman function is now given by \cite{birrell1984quantum}
    \begin{equation}
    \label{eq:wightman}
        W(\sx,\sx') = -\frac{1}{4\pi}\log\Bigg[\frac{(\epsilon+i(p(u)-p(u')) (\epsilon+i(v-v'))}{(\epsilon+i(p(u)-v')(\epsilon+i(v-p(u')))}\Bigg]
    \end{equation}
    For a static mirror located at the origin $x=0$ we have $p(u)=u$, and the result reduces to the well known fact that the Wightman function is the difference between the free-space Wightman function and its parity-reversed counterpart \cite{Hodgkinson:2013tsa}.
         \begin{figure}
        \centering
        \includegraphics[scale=0.65]{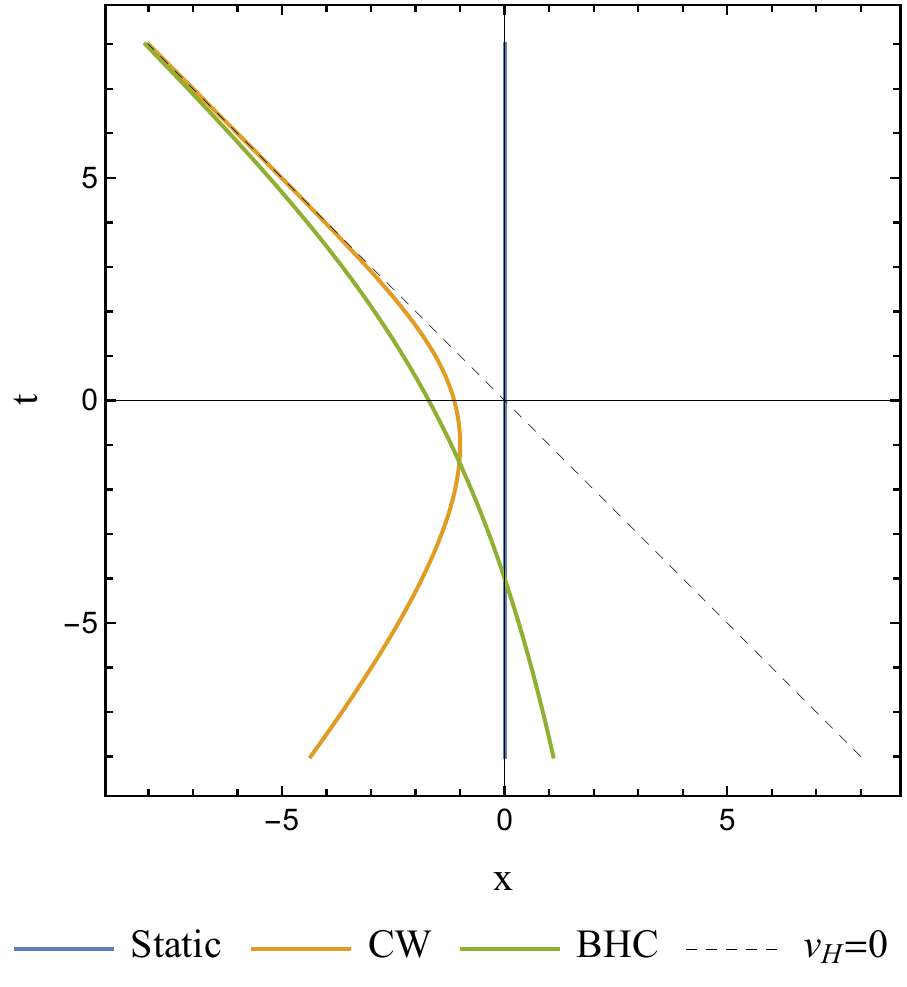}
        \caption{The various mirror trajectories considered in this paper are shown here, with $\kappa=1/2$ for CW and $\kappa=1/4$ and $v_H=0$ for BHC.}
        \label{fig: trajectories}
         \end{figure}      
   
    In this paper we will consider three ray-tracing functions, corresponding to three different trajectories \cite{Good2013mirror}:
    \begin{equation}
    \label{eq: raytracing}
        \begin{split}
        p_{0}(u) &= u\,,\\
        p_{1}(u) &= -\frac{1}{\kappa}e^{-\kappa u}\,,\\
        p_2(u) &= v_H - \frac{1}{\kappa} \textsc{W}\rr{e^{-\kappa(u-v_H)}} 
        \end{split}
    \end{equation}
    where $ \textsc{W}(x)$ is the Lambert $ \textsc{W}$ function \cite{Corless1996}.  The Lambert $ \textsc{W}$ function (sometimes known as \textit{product logarithm}) is not to be confused with Wightman function $W(\sx,\sx')$. The corresponding trajectories are shown in Figure~\ref{fig: trajectories}.

    The ray-tracing function $p_0$ describes a static mirror located at the origin. The function $p_1$ corresponds to a mirror that emits thermal radiation just like that of an eternal black hole, and is known as the
    \textit{Carlitz-Willey} (CW) trajectory \cite{CarlitzWilley1987reflection}. 
    We shall call the last ray-tracing function $p_2$ the \textit{black hole collapse} (BHC) trajectory, since it has been shown that there is one-to-one correspondence between the Bogoliubov coefficients for this moving mirror setup and the scenario of $(1+1)$ null shockwave collapse \cite{Good2016blackholemirror1,Good2016blackholemirror2,Good2016blackholemirror3}. In both cases
    the parameter $\kappa$ can be interpreted as some kind of acceleration parameter since both $p_1,p_2$ correspond to non-inertial motion of the mirror.

    \section{Results and discussions}
    \label{sec: results}
    We present the main findings in this section. The variable parameters will be denoted as follows: $(t_j,x_j)$ for the time and position coordinates of the peak of the Gaussian switching of detector $j$, $\Omega$ for the energy gap of the detectors, $d_A$ for the distance of detector $A$ from the mirror at $t=t_A$ and finally $\Delta x = x_B-x_A$ for the detector separation. The results will be presented in terms of the corresponding dimensionless variables, $\{t_A/\sigma,\, d_A/\sigma,\,\Omega_j\sigma,\,\Delta x/\sigma\}$.  We first describe general results for all mirror trajectories, followed by an analysis of the effects due to specific mirror motions.
    
    \subsection{Entanglement enhancement by a mirror}
    
    A mirror serves as a Dirichlet boundary condition for the quantum field along the mirror trajectory, i.e.
    \begin{equation}
    \label{eq:dirichlet}
        \hat\phi(z(t)) = 0 
    \end{equation}
    where $z(t)$ is the mirror trajectory parametrized by, say, Minkowski time $t$. Consequently the Wightman function must also vanish wherever one of the detector trajectories coincides with the mirror trajectory.
    Since the Wightman function gives a measure of correlations present in the field, we would expect to see a drop in the entanglement harvested when the mirror is approached. 
    
 \begin{figure}[tp]
    \includegraphics[scale=0.8]{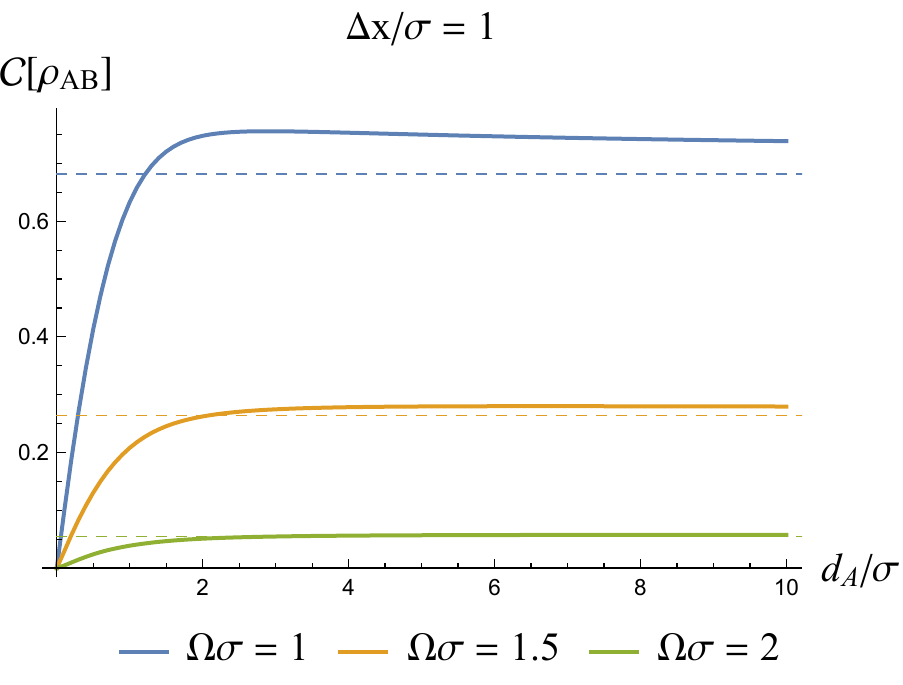}
    \includegraphics[scale=0.8]{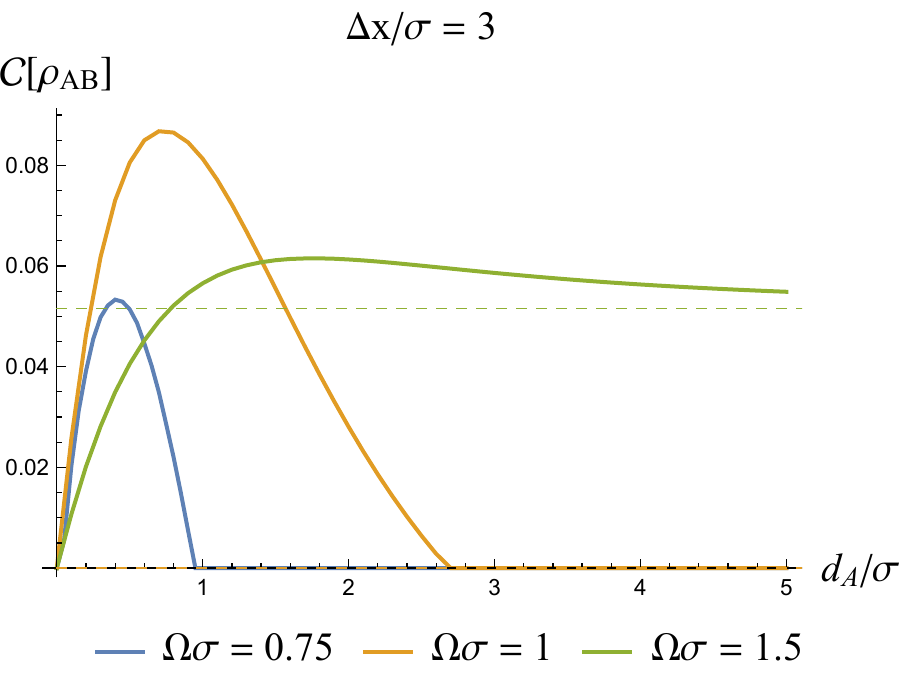}
    \caption{Concurrence as a function of distance from mirror $d_A$ with $\sigma = 1$ and for various energy gaps. The left and right plots are obtained for different detector separations $\Delta x/\sigma$ as indicated in the plots. The corresponding free space results are shown as dashed lines in each plot. We observe entanglement enhancement relative to that of free-space   in some regions. For $\Delta x/\sigma=3$ (right plot) and $\Omega\sigma=0.75,1$, the free space cases (computed by choosing $\Lambda = 10^{-12}$) have zero concurrence (dashed lines on $d_A/\sigma$ axis).}
    \label{fig: distfromStatic}
    \end{figure}  
    
    For a static mirror, we show in Figure~\ref{fig: distfromStatic} the concurrence $C[\rho_{AB}]$ against $d_A/\sigma$ plot for different values of the energy gap $\Omega\sigma$. Somewhat surprisingly, for fixed detector separation $\Delta x/\sigma$, we find that the presence of a mirror can actually {\it enhance} entanglement.
    
    Let us analyse the results in greater detail. Firstly, observe that for all parameter choices, the entanglement increases initially as the detectors are positioned further away from the mirror due to Dirichlet boundary condition in Eq.~\eqref{eq:dirichlet}. Far from the mirror, concurrence vanishes for large $\Delta x$ and small $\Omega\sigma$ (the blue and orange curves in the right figure). When the detector separation $\Delta x$ is decreased it becomes easier to harvest entanglement as we expect. Indeed, for very small $\Delta x$ (Figure~\ref{fig: distfromStatic}, left), the region of entanglement extraction is very large (possibly everywhere $d_A>0$). Conversely, for sufficiently large $\Delta x$ the concurrence vanishes and so entanglement cannot be extracted anywhere. 
    
    The more interesting observation from Figure~\ref{fig: distfromStatic} is that concurrence in the presence of a mirror can overtake the free-space result (dashed lines on the figure) at large enough $d_A/\sigma$ and small enough detector separation $\Delta x/\sigma$. As a representative example, consider $\Delta x/\sigma = 3, \Omega\sigma = 1$.  Noting that the free space case (dashed line) has zero concurrence, we see that entanglement harvesting would not have been possible at all if not for the presence of the mirror. Heuristically, this can be understood as a reflection effect in which information from one detector can reach the other detector after reflecting off the mirror. The trade off between this reflection effect and the vanishing of the Wightman function close to the mirror leads to a peak in the concurrence at some optimal $d_A$ away from the mirror. This qualitative behaviour is also present for the other mirror trajectories considered. The main feature displayed here is that of \textit{entanglement enhancement}: mirrors can amplify entanglement extraction relative to the free-space scenario. 
    Accelerating mirrors contain richer entanglement dynamics from the static one because the Wightman function is generally non-stationary. Furthermore, for finite switching width $\sigma$, in general there is some $t\in \left[-\frac{\sigma}{2},\frac{\sigma}{2}\right]$ in which $\hat\phi(z(t))\neq 0$, so naively we do not expect entanglement to completely vanish even if the detector coincides with the mirror at some $t\in\left [-\frac{\sigma}{2},\frac{\sigma}{2}\right]$. We will see below that the dynamics of entanglement is indeed quite non-trivial, as we explicitly demonstrate for the   CW trajectory.

    \subsection{Entanglement death near moving mirrors}
    
    \begin{figure}[tp]
    \centering
        \includegraphics[scale=0.51]{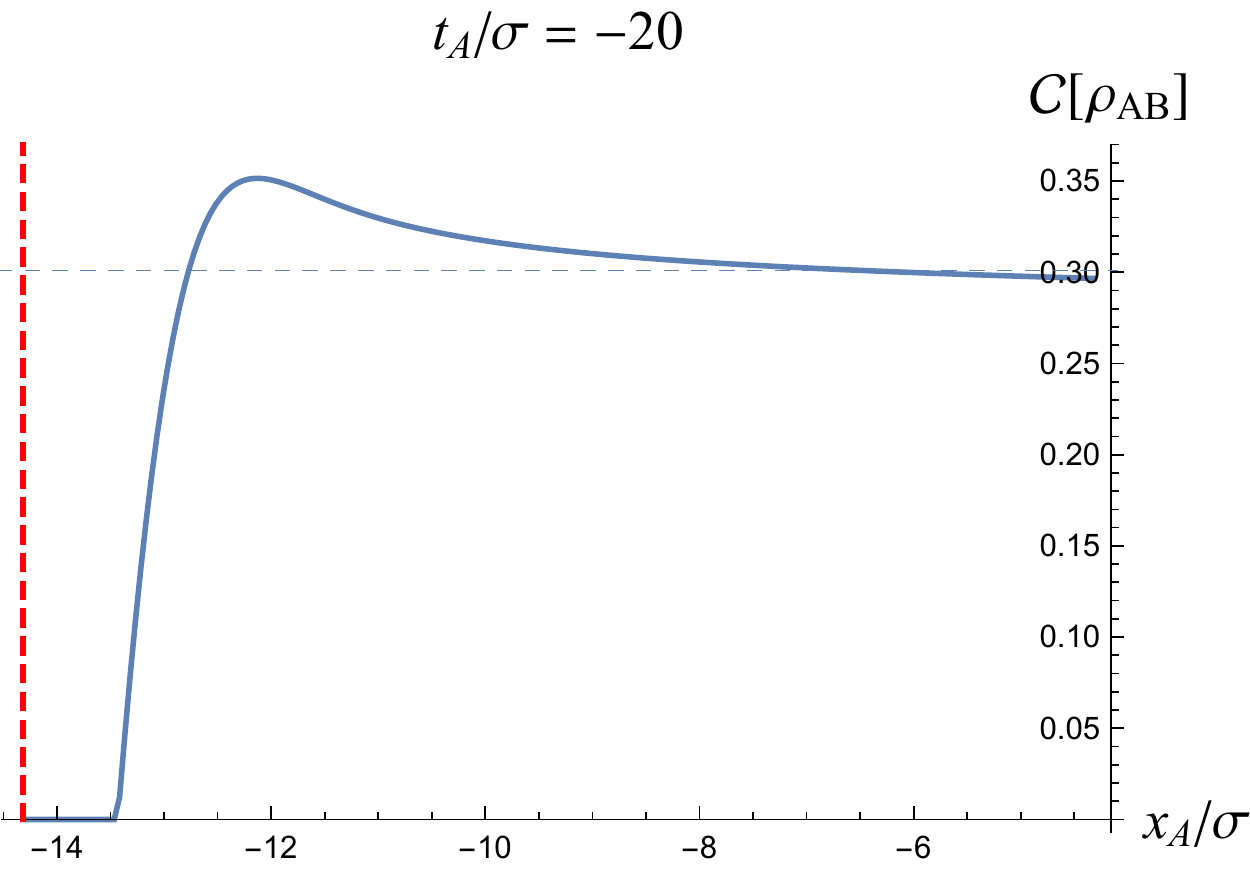} \quad
        \includegraphics[scale=0.51]{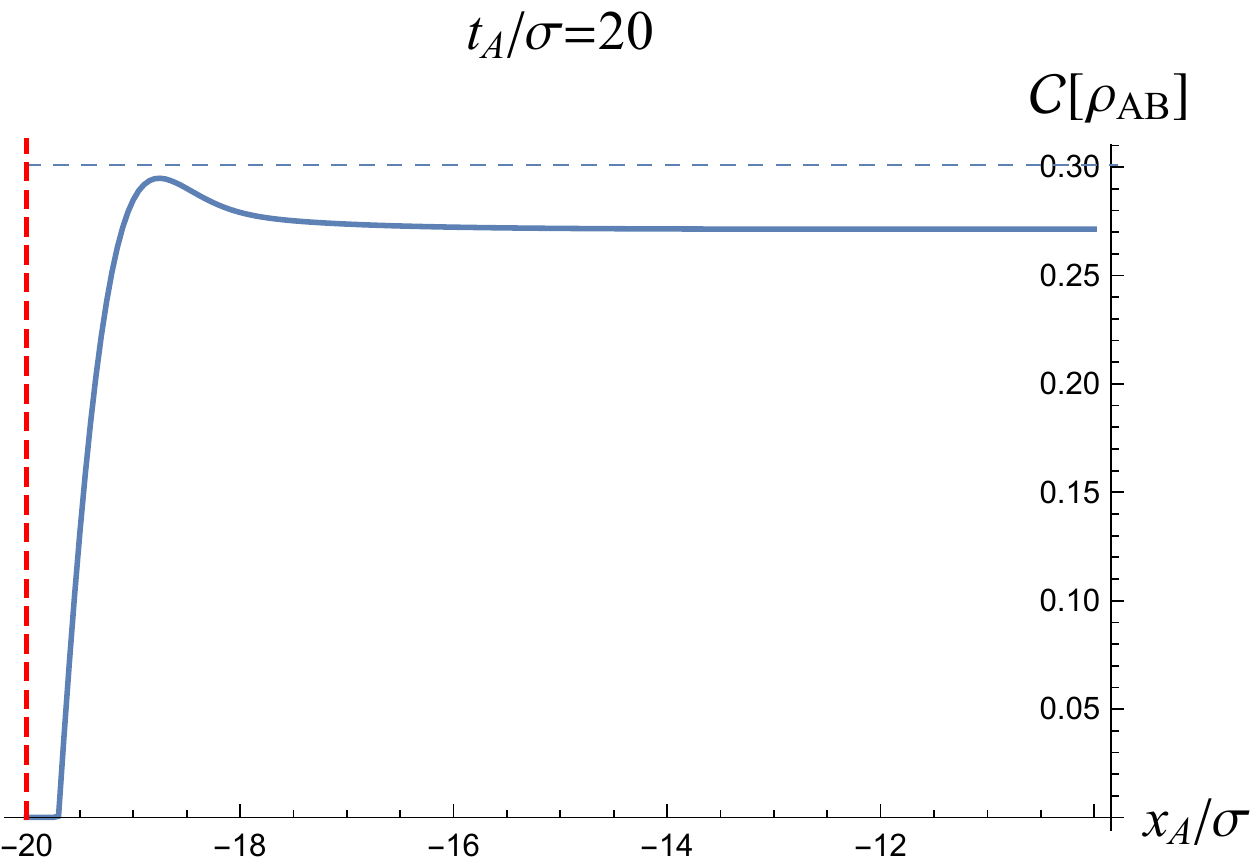}
        \caption{Concurrence as a function of position of detector A for fixed detector separations $\Delta x = 2\sigma$, with $\kappa= 0.5, \Omega=1,\sigma = 1$ for CW mirror (in red at time $t=t_A=t_B$ when the switching function is peaked). Note that $x_A/\sigma,\Omega\sigma, \kappa\sigma$ are dimensionless and numerically equal to $x_A,\kappa,\Omega$ since $\sigma=1$.
        \textbf{Left:} $t_A/\sigma=-20$. \textbf{Right:} $t_A/\sigma=20$. The entanglement extracted by the detectors are strictly lower than free space, and in both case there is a finite region near the mirror where entanglement extraction is impossible.}
        \label{fig: distfromCW}
    \end{figure}
    
\begin{figure}
    \centering
        \includegraphics[scale=0.5]{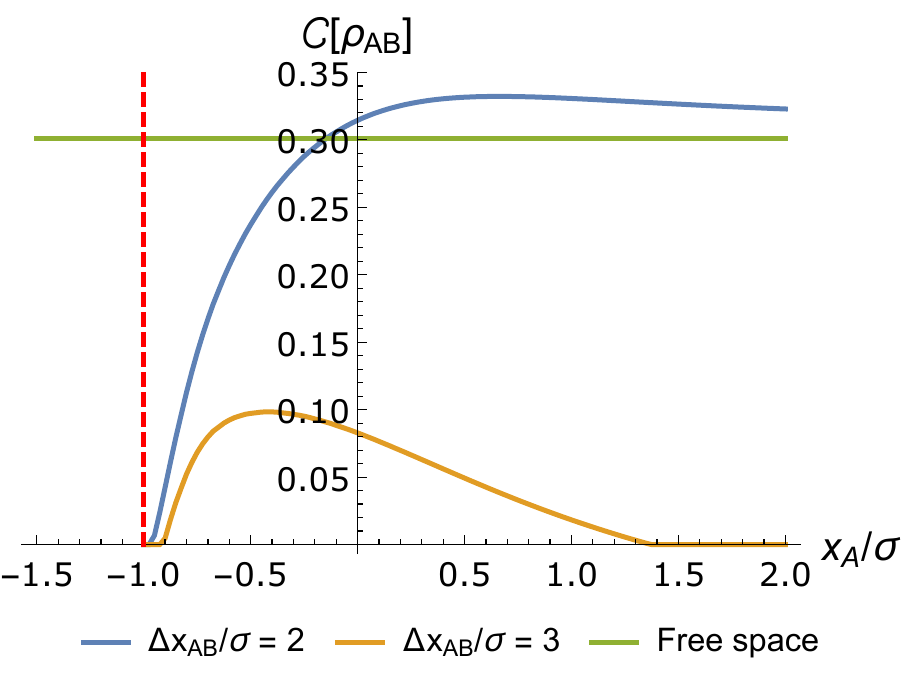}\quad 
        \includegraphics[scale=0.5]{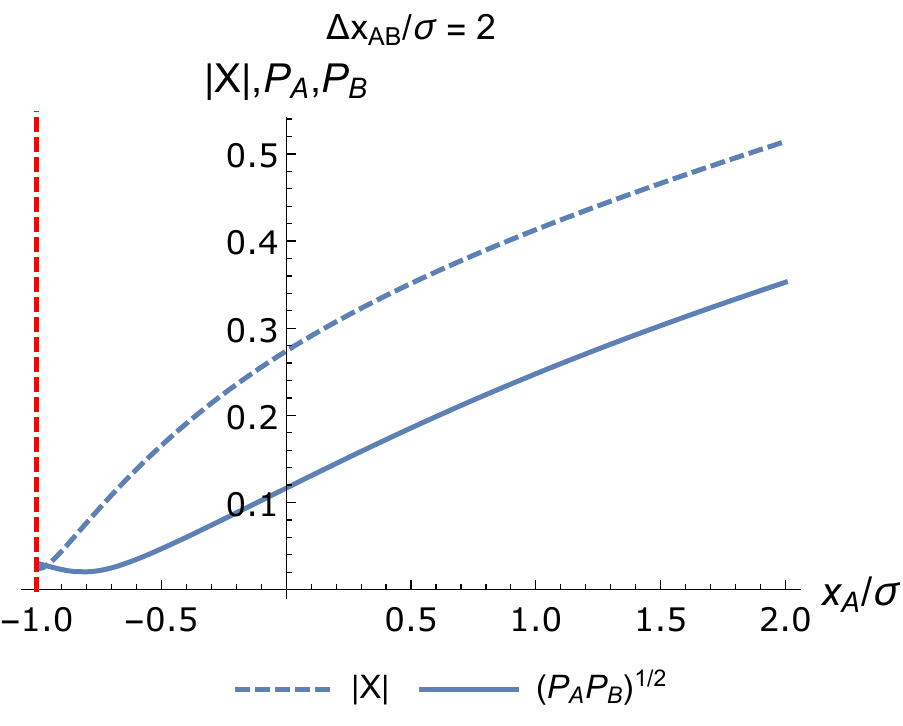}\quad \includegraphics[scale=0.5]{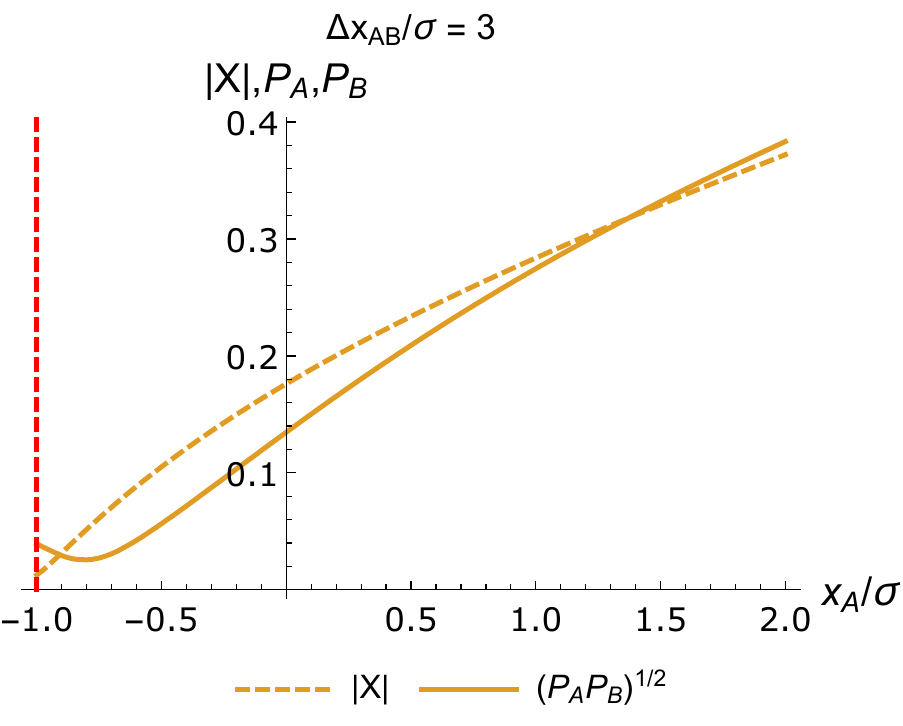}\quad
        \caption{\textbf{Left:} A plot of concurrence for the CW mirror (in red at time $t=-1$), as a function of the position of detector $A$ for various fixed detector separations
        $\Delta x$, with $\kappa= 0.5, \Omega=1,\sigma = 1$.  \textbf{Middle:}
        The nonlocal term $|X|$ and local noise $\sqrt{P_AP_B}$ terms for the $\Delta x = 2\sigma$ trajectory in
        the left figure. \textbf{Right:} The nonlocal term $|X|$ and local noise $\sqrt{P_AP_B}$ terms for the $\Delta x = 3\sigma$ trajectory in    the left figure.   The small region where the two curves intersect give the small entanglement enhancement region. Note the small zones of entanglement death near the mirror on the left plot for both cases; these appear in the other two plots where we see that $|X| < \sqrt{P_AP_B}$ very close to the mirror.}
        \label{fig: CWcompete1}
    \end{figure}

    The results for the CW mirror are shown in Figures~\ref{fig: distfromCW} and~\ref{fig: CWcompete1}. 
    Such mirrors are known to have a constant flux of Hawking radiation. We focus on one particular choice of $\kappa\sigma = 0.5,\Omega\sigma=1$ which captures all the qualitative features we hope to highlight.
    
    The CW mirror scenario contains more interesting physics compared to its static counterpart. In contrast to the static mirror case, where entanglement vanishes strictly at the mirror due to the boundary condition (cf. Figure~\ref{fig: distfromStatic}),
    we see from Figure~\ref{fig: distfromCW} that there is a small finite region of \textit{entanglement death} near the mirror. This is reminiscent of the situation when detectors are placed too close to a black hole event horizon \cite{Henderson:2017yuv}, but the physical origin is different since there is no black hole in our case. For the black hole the origin of entanglement death is due to a redshift factor diminishing the non-local correlations relative to the local noise terms. In the present case, superficially there is a (nonlinear) competition between the local noise term $\sqrt{P_AP_B}$ and the nonlocal term $|X|$ due to the logarithmic behaviour of the Wightman function since they grow at different pace with distance from mirror at some fixed time $t$, as shown in the left plot of Figure~\ref{fig: CWcompete1}.
    
    This result is interesting because the CW mirror trajectory $z(t)$ has a future horizon \cite{CarlitzWilley1987reflection}. This is the line $v_H = 0$ in Figure~\ref{fig: trajectories}. The trajectory is asymptotically null in the infinite past and future, i.e.
    \begin{equation}
        \frac{\dd z(t)}{\dd t} = \frac{2  \textsc{W}\left(e^{-2 \kappa t}\right)}{ \textsc{W}\left(e^{-2 \kappa t}\right)+1}-1\Longrightarrow \lim_{t\to\pm\infty}\frac{\dd z}{\dd t} = \pm 1\,.
    \end{equation}
    Similar to black hole event horizon, the mirror's accelerating horizon prevents some modes at past null infinity $\skri^-$ from reaching (being ray-traced to) null infinity $\skri^+$. This suggests that the analogy/mapping between accelerating mirror and black hole spacetimes can be understood as mapping horizons in both spacetimes. This is appropriate for the CW mirror, since the corresponding black hole \cite{Good2013mirror} is eternal black hole with two horizons conventionally denoted $\mathcal{H}^\pm$.
    
    CW mirrors can also enhance entanglement. The size of the region of entanglement enhancement depends on the relative separation of the two detectors, as the left plot of Figure~\ref{fig: CWcompete1} shows. Meanwhile, Figure~\ref{fig: distfromCW} demonstrates that the entanglement structure of this mirror spacetime is not time-symmetric even though the thermal spectrum is time independent; unlike the radiation spectrum, concurrence is sensitive to the non-stationary nature of the spacetime.  At ``early'' times (left plot), in addition to a finite region of entanglement death we observe a finite region of entanglement enhancement relative to the free space result. However, at ``late'' times (right plot) the size of the entanglement death zone decreases and the harvesting zone 
    always yields an amount of concurrence strictly less than free-space limit. 
    This is likely due to the fact that at $t=20$ the mirror is already at ultrarelativistic speed. From this we see that generically at late times an accelerating mirror inhibits entanglement compared to early times.

    To summarize, we see that the main effect a non-trivial mirror trajectory has on entanglement harvesting is the generic presence of  an \textit{entanglement death zone} near the mirror. The strip where this occurs may increase or decrease in size depending on the proximity of the detectors, as shown in Figure~\ref{fig: distfromCW}. 
    The fact that $\concur$ is sensitive to non-stationarity of the mirror also suggests that entanglement is not correlated directly with a thermal flux of radiation, since a CW mirror models a constant flux of Hawking radiation but $\concur$ is clearly dependent on when the detector is switched on.

    \subsection{Effect of different trajectories}
    \label{section: blackhole}
    
     \begin{figure}[tp]
     \centering
        \includegraphics[scale=0.5]{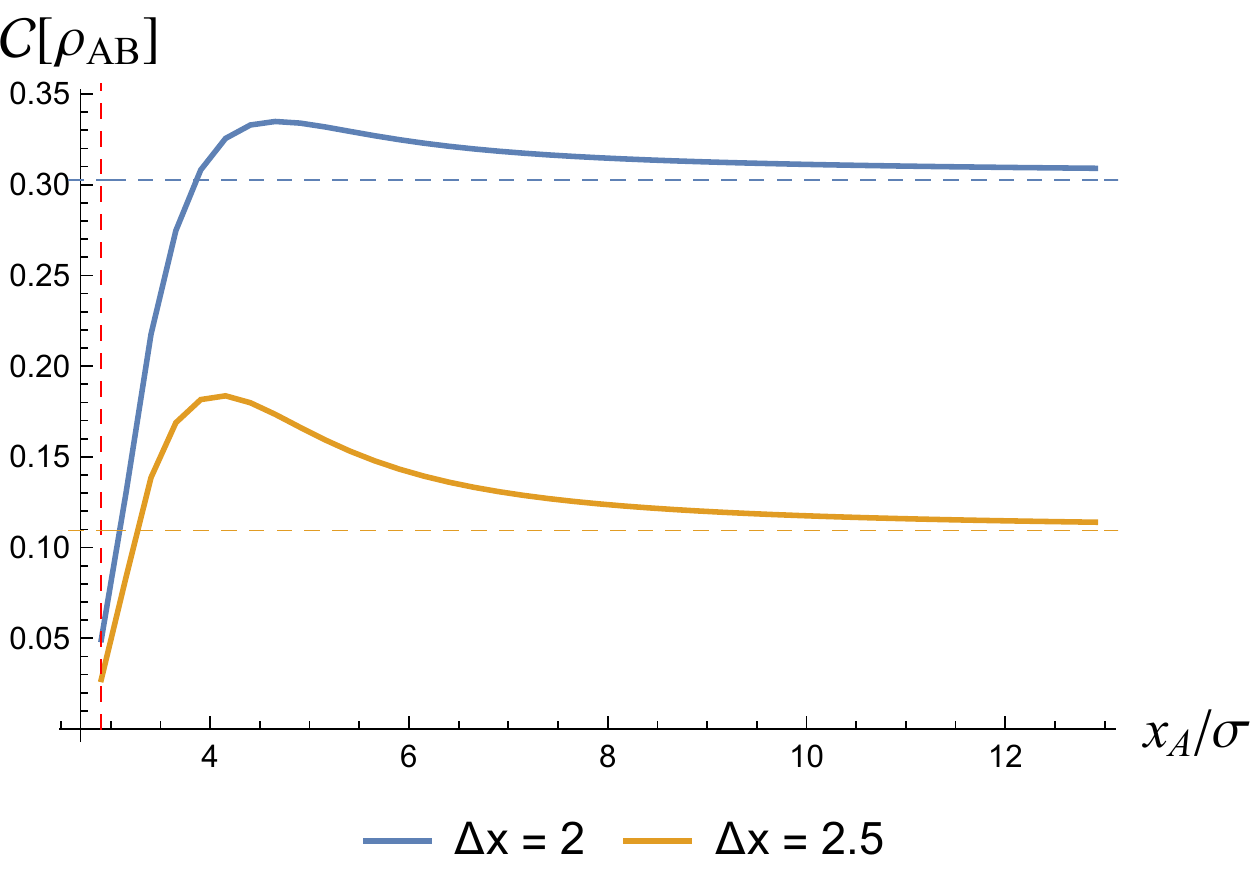}\quad
        \includegraphics[scale=0.65]{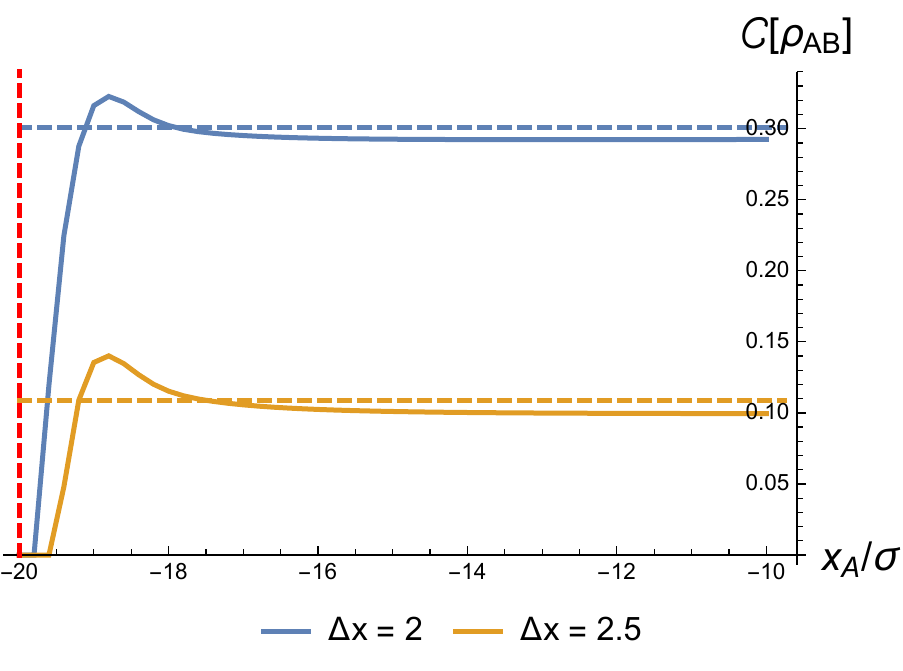}
        \caption{Concurrence for a   BHC mirror (in red at time $t$ when the detector is temporally peaked) for various fixed detector separations as a function of position of detector $A$ with $\kappa= 0.25, \Omega=1,\sigma = 1$. From left to right: \textbf{Left} $v_H=0$, the ``earlier''time $t=-20$. \textbf{Right} $v_H = 0$, the ``late'' time $t=20$.}
        \label{fig: distfromBHC}
    \end{figure}
    
    While the Carlitz-Willey trajectory models black hole radiation, it does not model black hole collapse because particle creation due to an accelerating mirror following a CW trajectory is \textit{always} thermal \cite{Good2016blackholemirror2}. This is understandable since the CW trajectory has an exact one-to-one correspondence with an eternal black hole \cite{Good2018Evanescent}. Modelling black hole collapse on the other hand requires a mirror trajectory in which  particle creation is only thermal at late times. In (1+1) dimensions, the analysis is surprisingly manageable and  is provided by the mirror trajectory \cite{Good2016blackholemirror1,Good2016blackholemirror2}
    \begin{equation}
    \label{eq: bhc}
        z(t) = v_H - t - \frac{ \textsc{W}(2e^{2\kappa(v_H-t)})}{2\kappa}
    \end{equation}
    where to model collapse to a black hole of mass $M$ the identification $\kappa=1/4M$ is made and the horizon is given by the future light cone of the point with double-null coordinate $(v_H,v_0)$ where $v_H = v_0-4M$ \cite{Good2016blackholemirror2}. 
    This trajectory has zero velocity and acceleration in the asymptotic past at infinity and models the shock-wave collapse of a null shell at $v=v_0$, i.e. the spacetime that is flat for $v<v_0$ and Schwarzschild for $v>v_0$. We shall call this trajectory the ``black hole collapse'' (BHC) trajectory. 

Since both CW and BHC trajectories are asymptotically null in the future, we might expect that their differences must arise at early and intermediate times. We note that unlike the CW trajectory, the BHC mirror is at $x=\infty$ when $t=-\infty$ whereas the CW mirror is at $x=-\infty$ (see Figure~\ref{fig: trajectories}). As such, for the BHC trajectory, in principle the mirror will cross the detectors at some time; in this case, we imagine that the detectors have Gaussian switching that effectively makes the detector active only when the mirror is on the left side of both detectors  and imposes the boundary condition so that the field vanishes on the left side of the mirror. 

We are primarily interested in the generic behaviour of entanglement between the detectors as compared to the static and CW mirror cases. First, we obtained expected dependence on the detector separation $\Delta x/\sigma$ (not illustrated) generic to all three mirror trajectories: for small $\Delta x/\sigma$, we find that we can extract entanglement from the vacuum very far away from the mirror; for large $\Delta x/\sigma$, no entanglement can be extracted at all for any $d_A/\sigma$. 
    
The situation becomes more interesting  when we choose to turn on the detectors at different times, to see what ``early'' or ``late'' times do to the concurrence. Unlike the CW trajectory, the BHC mirror trajectory has no turning point, i.e. it always moves in the negative $x$ direction, and so it is not obvious what will happen here. The results are illustrated in Figure~\ref{fig: distfromBHC}, and turn out to be qualitatively similar to the CW trajectory. This is  particularly clear upon comparing the plots in Figure~\ref{fig: distfromBHC}. We see a degradation of entanglement near the mirror at early (left) times that widens into a death zone at late (right) times, which is likely due to the differing magnitudes of the mirror velocity.  

    
 Last but not least, we note that the overall entanglement inhibition at late times when the mirror trajectory is almost null, as well as the entanglement death near the mirror are generic features of accelerating mirror spacetimes. Since for BHC trajectory we have
 \begin{equation}
     \frac{\dd z(t)}{\dd t} = \frac{ \textsc{W}\left(2 e^{2 \kappa (v_H-t)}\right)}{ \textsc{W}\left(2 e^{2 \kappa (v_H-t)}\right)+1}-1
 \end{equation}    
 which gives $\lim_{t\to-\infty}\dot z(t) = 0$ and $\lim_{t\to\infty}\dot{z}=1$, the mirror only has one accelerating (future) horizon, which is similar to a collapsing Vaidja-type spacetime (which also only has one apparent horizon asymptotically approaching future event horizon $\mathcal{H}^+$). This similarity is not a coincidence.
 
 Overall, our study on entanglement extraction by two \textit{localized} detectors in these mirror spacetimes provide further indication that we should take seriously that one can establish correspondence between the mirror spacetimes and black hole spacetimes as argued in \cite{Good2013mirror,Good2016blackholemirror1,Good2016blackholemirror2,Good2016blackholemirror3} where the correspondence is made based on nonlocal, non-observable quantities, namely that their Bogoliubov coefficients agree.

  \subsection{Detector Response and Concurrence in the presence of a boundary}    
  \label{sec: distancedep}
       \begin{figure}[tp]
       \centering
        \includegraphics[scale=0.5]{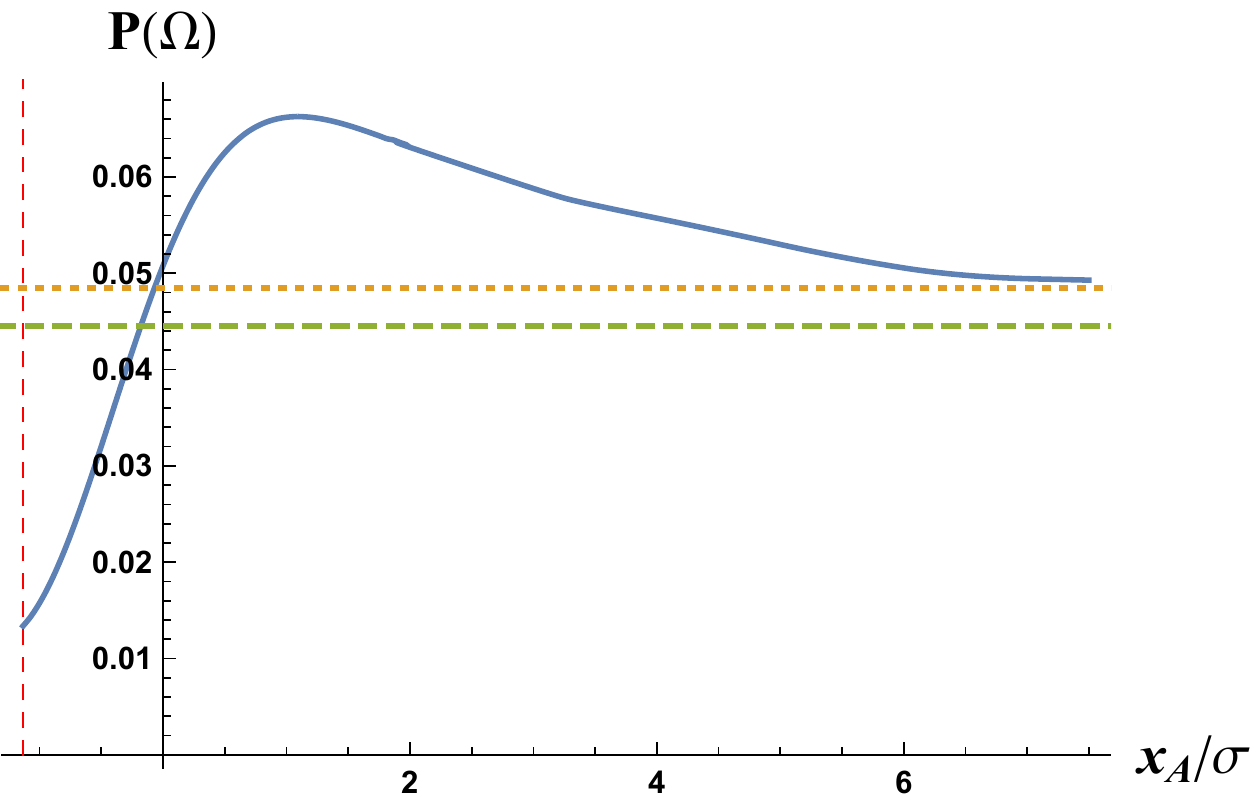}\quad
        \includegraphics[scale=0.5]{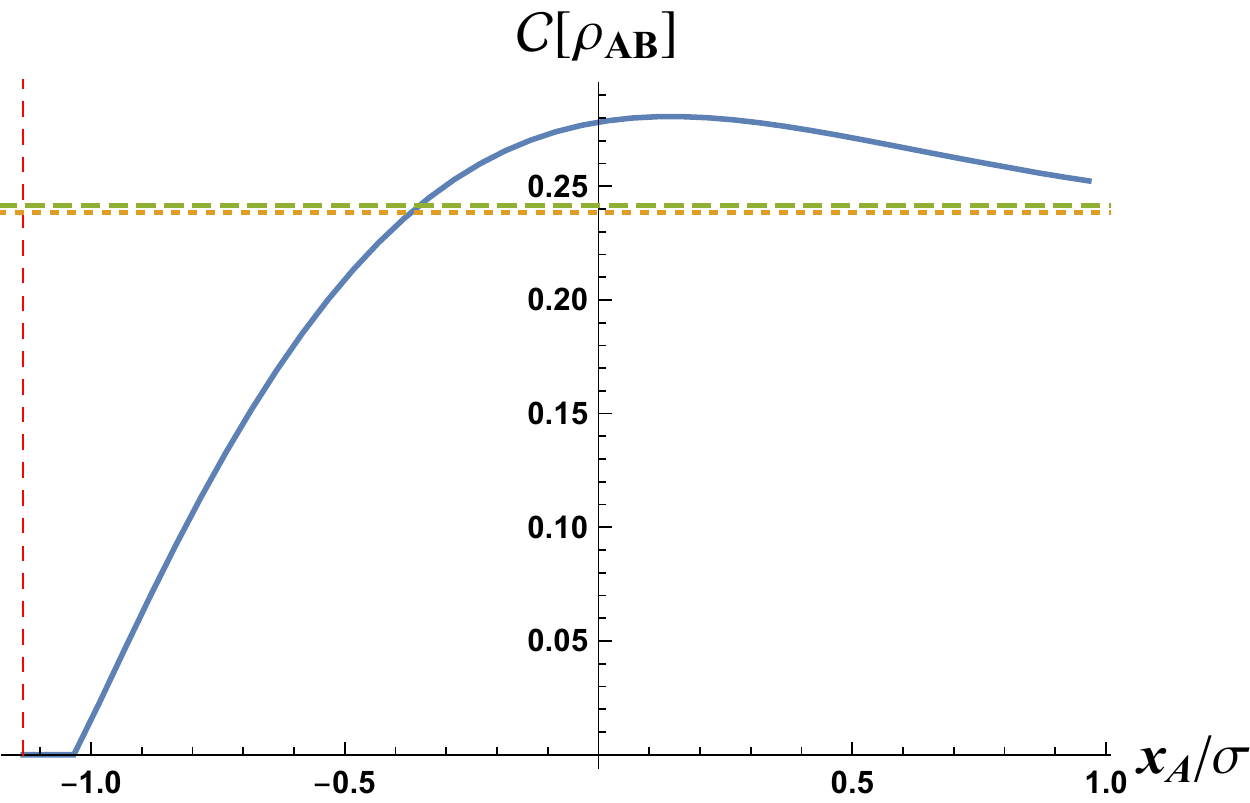}
        \caption{Derivative coupling results. \textbf{Left}: The use of derivative coupling removes the IR ambiguity in the free-space $P(\Omega)$ (dashed, green). The probability of excitation in the CW mirror spacetime now remains bounded at large $x_A/\sigma$. It asymptotes to a value (dotted, yellow) slightly higher than the free-space result. \textbf{Right}: The concurrence against $x_A/\sigma$ plot shows the same qualitative behaviour as linear coupling. Namely, we observe a region of entanglement enhancement and a region of entanglement death close to the mirror. The concurrence will asymptote to a value (dotted, yellow) slightly below the free-space result (dashed, green). In this plot, $\kappa\sigma = 0.5$, $\Omega\sigma = 1$, $t_A/\sigma = 0$ and $\Delta x/\sigma = 1$.}
        \label{fig: DerPlots}
    \end{figure}
 
 Upon closer scrutiny, two subtleties in the above sections may cause one to question the validity of the results. The first is the ambiguity in the free-space concurrence results due to the infrared cut-off and the second is the spurious effect of an unbounded growth in the excitation probability of a detector in the mirror spacetime (see Figure~\ref{fig: CWcompete1}). We include an appendix~\ref{sec: Appendix} detailing the origin of the blow up in $P(\Omega)$ and how $|X|$ compensates this to result in sensible results for the concurrence in mirror spacetimes. We also show there how the free-space concurrence is independent of the infrared cut-off. 
 
 However, these two subtleties can be bypassed altogether by considering a derivative type coupling (see for example \cite{Ju_rez_Aubry_2014}) between the detector and the field. In this subsection, we employ this alternative coupling, and show that in the absence of the above two effects, the qualitative results obtained in the previous sections still hold.
 
 Instead of the linear coupling between the detector and the field operator $\hat{\phi}$ in Eq.~\ref{eq: int}, we can replace $\hat{\phi}(\sx(t))$ with its proper time derivative to obtain the derivative coupling,
 \begin{equation*}
        \tilde{H}_I^j = \lambda\chi_j(\tau)\hat\mu_j(\tau)\otimes u^\mu\nabla_\mu\hat\phi(\sx_j(\tau))\,. 
 \end{equation*}
 For detectors static in the $(t,x)$ coordinates, we have $t=\tau$ and the proper time derivative reduces to partial derivative $\partial_t$. 
 
 As shown in \cite{Ju_rez_Aubry_2014}, in addition to removing the dependence of the excitation probability on the infrared cut-off, this coupling also results in an expression for the probability that looks more similar to the $(3+1)$D case. In the current case, we can see this by the following: the expressions for $P(\Omega)$ and $X$ given in Sec.~\ref{sec: setup} can be easily modified to accommodate the change in coupling by making the replacement $W(\sx_{j_1}(t),\sx_{j_2}(t'))\rightarrow A(\sx_{j_1}(t),\sx_{j_2}(t')) \equiv \partial_t\partial_{t'}W(\sx_{j_1}(t),\sx_{j_2}(t'))$ whenever it appears. This replacement yields a well-defined expression\footnote{In particular, since for identical static detectors we considered here the negativity $\mathcal{N}$ is given by \cite{Henderson:2017yuv,Allison2017quad}
 \begin{align}
     \mathcal{N} = \max\left\{ 0,\sqrt{|X|^2-\rr{\frac{P_A-P_B}{2}}^2}-\frac{P_A+P_B}{2}\right\}\,,
 \end{align}
 this expression is well-defined if the density matrix is. For the same reason concurrence $\concur$ will also be well-defined.
 } for the reduced detector density matrix $\hat\rho_{AB}$ because the distributional singularities of the derivative Wightman $A$ do not coincide with the finite discontinuities of the Heaviside function appearing in Eq.~\eqref{eq: nonlocal}.
 
 In particular, we have 
 \begin{align}
     A_f(\sx,\sx') &= -\frac{1}{4\pi}\Bigg(\frac{1}{\Delta u-i\epsilon}+\frac{1}{\Delta v -i\epsilon}\Bigg)\,,\\
     A_m(\sx,\sx') &=-\frac{1}{4\pi}\Bigg(\frac{p'(u)p'(u')}{\big[p(u)-p(u')-i\epsilon\big]^2}+\frac{1}{\big[v-v'-i\epsilon\big]^2}\\ \nonumber
     &\hspace{1.25cm} -\frac{p'(u')}{\big[v-p(u')-i\epsilon\big]^2}-\frac{p'(u)}{\big[p(u)-v'-i\epsilon\big]^2}\Bigg)
 \end{align}
 respectively for free space and mirror spacetimes. These replace the Wightman functions in \eqref{eq: freeW} and  \eqref{eq:wightman}. Since $A_f$ does not require an IR regulator to be well-behaved, the IR ambiguity in $P(\Omega)$ and $|X|$ is removed. Furthermore, the similarity between $A_f$ and the $(3+1)$D Wightman function (c.f. Appendix~\ref{sec: Appendix}) also indicates that the results for $P(\Omega)$ and $|X|$ in $(3+1)$D using linear coupling will be similar to that obtained using derivative coupling in ($1+1)$D. In fact, for the free space scenario   the Wightman function $A_f(\sx,\sx')$
 in $(1+1)$ dimensions only differs from the linear coupling Wightman function $W_f(\sx,\sx')$ in $(3+1)$D by a constant factor of 2, so the physics is practically identical.
    
    In Figure~\ref{fig: DerPlots}, we show the results obtained using derivative couplings between the detectors and the field. From the $P(\Omega)$ plot, we see that the probability in the CW mirror spacetime remains bounded at large $d_A$ rather than blowing up as in Figure~\ref{fig: CWcompete1}. We note also that the free-space value in this case was computed without the need for choosing an IR cut-off. In addition, we see that derivative coupling results in the same qualitative findings as the previous subsections: there is a region of entanglement enhancement over the free-space result and entanglement death near the mirror. 
    
    The derivative coupling scenario reinforces the fact that entanglement dynamics between detectors in mirror spacetimes can be understood as intrinsically the physics of horizons: even though the time-evolved reduced density matrix depends strongly on the type of coupling used, we see that entanglement itself is qualitatively robust and captures the essential physics induced by presence of horizons. Our study complements the correspondence previously established in \cite{Good2013mirror,Good2016blackholemirror1,Good2016blackholemirror3,Good2016blackholemirror2,Good2018Evanescent} using nonlocal/non-observable quantities such as Bogoliubov coefficients and expectation value of stress-energy tensor, by establishing qualitative correspondence via local measurements using detectors.

    \section{Conclusion}
    \label{sec: conclusion}

    We have performed in $(1+1)$ dimensions the first investigation of entanglement harvesting between two detectors in the presence of mirrors using both linear and derivative couplings between the detectors and the quantum field. We looked at both static and  non-inertial trajectories, focusing on the CW and BHC trajectories for the latter because of their respective correspondence to eternal black holes and black hole formation. We find that a similar correspondence exists for their entanglement structures, in particular the phenomenon of an entanglement death zone, similar to that recently found for a black hole event horizon \cite{Henderson:2017yuv}, which can be traced to the fact that both sides of the correspondence have analogous horizons. Physically, our results provide a theoretical prediction of what to expect of the entanglement detection in the presence of the DCE.
    
    
    Finally, and perhaps most interestingly, we also find that  mirrors can enhance entanglement: entanglement harvested between two detectors can be greater in the presence of a mirror as compared to free space. Qualitatively this enhancement is largely trajectory-independent, and is thus attributable to the inherent presence of a Dirichlet boundary condition.  Quantitatively there are distinctions between different mirror trajectories, as a comparison between Figures~\ref{fig: distfromStatic}  and~\ref{fig: distfromCW} indicates. We focused on three types of mirror trajectories that have the simplicity of having smooth ray-tracing functions $p_j(u)$ that are regular for all $u\in \R$. Important trajectories such as a mirror with constant uniform acceleration have ray-tracing functions that are not defined for all $u$, indicating that only some modes at future null infinity $\mathcal{I}^+$ can be `ray-traced' to modes at past null infinity $\mathcal{I}^-$. We have also not investigated the case of piecewise mirror trajectories,  such as a mirror that only accelerates at $t=0$ and static at $t\leq 0$, which are often used in analyses involving Bogoliubov transformations.  We defer this for future work. 
    
    \acknowledgments
    This work was supported in part by the Natural Sciences Engineering Research Council. Erickson Tjoa was supported by Mike-Ophelia Lazaridis Fellowship during this work. The authors thank Eduardo Mart\'in-Mart\'inez and Robie A. Hennigar for helpful discussions.

    \appendix
    \section{IR cut-off in free-space and unbounded transition probability in mirror spacetimes}
    \label{sec: Appendix}
    One peculiarity we found in this work is that for linearly coupled detector in $(1+1)$ dimensions, there seems to be a slow but unbounded growth of the transition probability $P(\Omega)$ as a function of distance from the boundary, which we denote by $d$. In particular, this means that in $(1+1)$ dimensions the static mirror does not smoothly recover the free space limit as $d\to\infty$, which is in contrast to the expectation that a detector does not sense nonlocal differences and in particular whether there is a boundary far away. This may influence the reliability of entanglement measures  that inherently depend on $|X|$ and $P(\Omega)$. Although the problem of particle detectors in half-space has been previously considered \cite{davies:1989boundary,suzuki:1997boundary,Brown:2015yma},  the growth of $P(\Omega)$ as a function of $d$ was not clarified. We will compare two scenarios for static Dirichlet boundary conditions in (1+1) and (3+1) dimensions. We will show that the peculiarity is dimension dependent and not generically true in the presence of a boundary.
    
\begin{figure}[htp]
        \centering
        \includegraphics[scale=1]{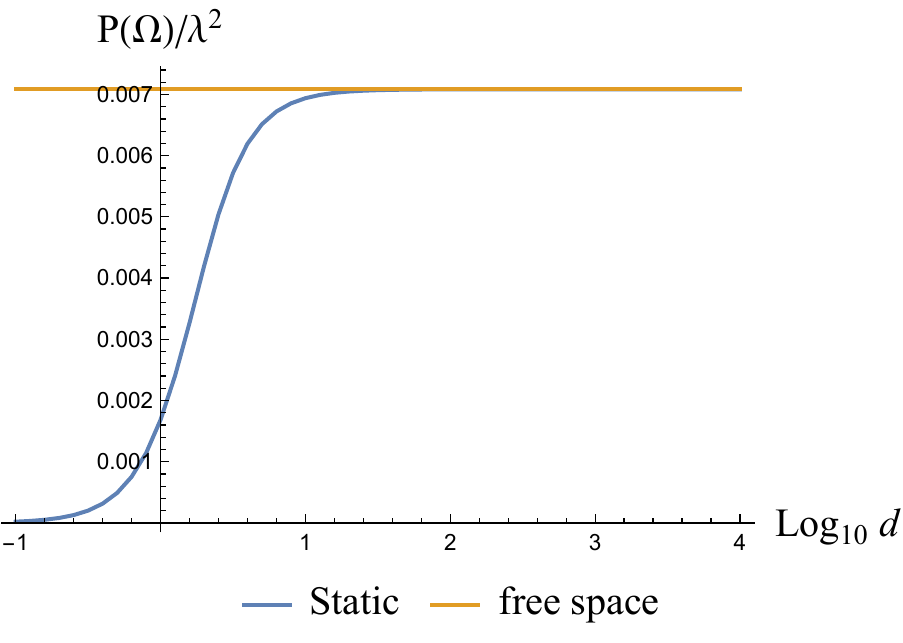}
        \caption{Transition probability for Gaussian switching $\sigma=1$ and $\Omega=1$ ($\Omega \sigma=1$) in $(3+1)D$. Near the mirror $P(\Omega)$ decays quickly, and far from the mirror the contribution from the image term vanishes quickly. In this figure the asymptotic behaviour is reached at $d/\sigma\sim 10^2$.}
        \label{fig: scaling3d}
    \end{figure}

    In $(3+1)$ dimensions, the Wightman function for a Dirichlet boundary on the $yz$-plane such that a detector is located at $(d,0,0)$ where $d>0$ is given by the image sum
    \begin{equation}
    \begin{split}
        W(\sx,\sx') = -\frac{1}{4\pi^2}\left[\frac{1}{(\Delta \tau-i\epsilon)^2}-\frac{1}{(\Delta \tau-i\epsilon)^2-4d^2}\right]
    \end{split}
    \end{equation}
    where the second term is the image term that  gives the distance dependence $d$. Evaluating the expression for $P(\Omega)$ in Eq.~\eqref{eq: probability} for this Wightman function and the usual Gaussian switching in Eq.~\eqref{eq: gaussswitch}, we will get two contributions to the integral:
    \begin{equation}
        P(\Omega) = P_{\text{free}}(\Omega)+P_{\text{image}}(\Omega) 
    \end{equation}
where  
  \begin{equation}
    \begin{split}
        P_{\text{free}}(\Omega) &= - \frac{\sigma}{\sqrt{16\pi^3}}  \int_{-\infty}^\infty\dd y \frac{e^{-\frac{y^2}{4\sigma^2}}e^{-i\Omega y}}{(y-i\epsilon)^2}\,,\\
        P_{\text{image}}(\Omega) &= \frac{\sigma}{\sqrt{16\pi^3}}  \int_{-\infty}^\infty\dd y \frac{e^{-\frac{y^2}{4\sigma^2}}e^{-i\Omega y}}{(y-i\epsilon)^2-4d^2}\,.
    \end{split}
    \end{equation}
    The first term is just the finite-time response of a detector in free space and hence is independent of $d$. The image term  \cite{padmanabhan:1996finite} is known to be finite even in free space
    for finite width $\sigma>0$. We are interested in the behaviour of the integral of the image term as $d$ increases, which we show in Figure~\ref{fig: scaling3d}. This is in fact not hard to see from the form of the Wightman function: as $d\to \infty$, the image term becomes more and more negligible because the finite width Gaussian window suppresses contributions at large $u$. Hence, in (3+1) dimensions the phenomenology of inertial (static) detectors in Minkowski half-space has the expected free-space limit  as $d\to \infty$.
    
    
\begin{figure}
        \centering
        \includegraphics[scale=1]{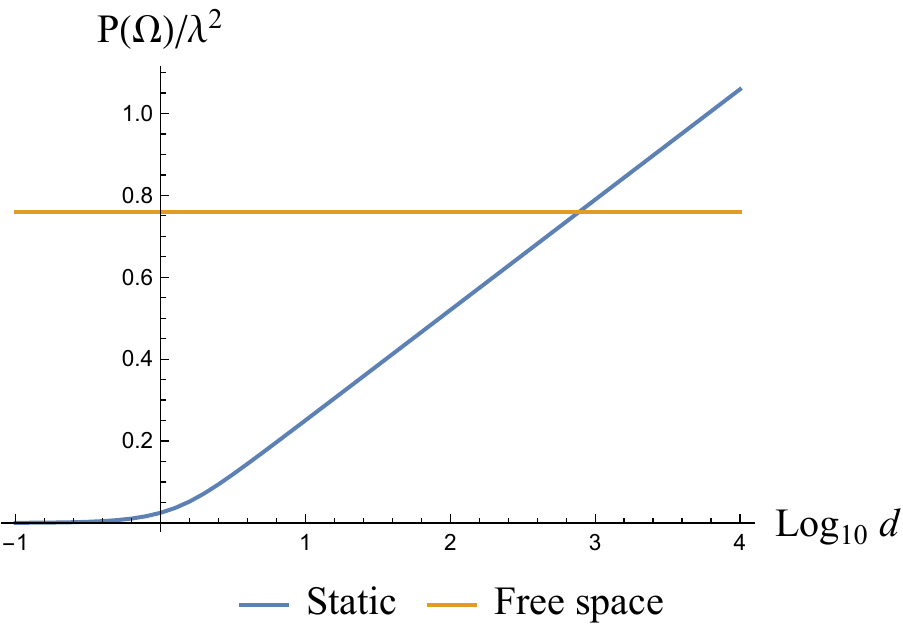}
        \caption{Transition probability for Gaussian switching $\sigma=1$ and $\Omega=1$ ($\Omega \sigma=1$) in $(1+1)$ dimensions. Near the mirror $P(\Omega)$ still decays quickly, but far from the mirror the contribution from the image term dominates and will exceed the free space case for any choice of IR cutoff $\Lambda$.}
        \label{fig: scaling1d}
    \end{figure}

    This, however, is not true in (1+1) dimensions because the Wightman function has a logarithmic instead of power law dependence on $d$. For a static mirror, the expression is instead 
    \begin{equation}
    \begin{split}
    P_{\text{free}}(\Omega) &= -\frac{\sigma}{\sqrt{16 \pi}}\int_{-\infty}^\infty \dd y\, e^{-i\Omega y}e^{-\frac{y^2}{4\sigma^2}}  \log\left[\Lambda^2(\epsilon+iy)^2\right]\,, \\
     P_{\text{image}}(\Omega) &=  -\frac{\sigma}{\sqrt{16 \pi}}\int_{-\infty}^\infty \dd y\, e^{-i\Omega y}e^{-\frac{y^2}{4\sigma^2}}  \log\left[\frac{1}{\Lambda^2\left((\epsilon+iy)^2+4d^2\right)}\right]\,,\\
    \Longrightarrow  P(\Omega) &=  -\frac{\sigma}{\sqrt{16 \pi}}\int_{-\infty}^\infty \dd y\, e^{-i\Omega y}e^{-\frac{y^2}{4\sigma^2}}  \log\left[\frac{(\epsilon+iy)^2}{(\epsilon+iy)^2+4d^2}\right]\,,
    \label{eq:totalprobmirror1D}
    \end{split}
    \end{equation}  
where we have made the free-space IR cutoff $\Lambda$ explicit for comparison. The branch of the logarithm is chosen to match that of Eq.~\eqref{eq: freeW}. Recall that in (1+1) dimensions, the free-space  Wightman function is infrared divergent. The common regularisation scheme employed in the literature is the addition of an IR cutoff, which can be thought of (for example) as the length scale of an optical fibre \cite{Pozas2015,Eduardo2015firewall}. Numerical results have been shown to be reliable as long as $\Lambda$ is chosen to be significantly smaller than the lowest frequency scale in the problem (i.e. the sensitivity to the choice of cutoff is very small). In the mirror spacetime, we note that the computation of $P(\Omega)$ does not require an IR cut-off as we see from the last expression in Eq.~\eqref{eq:totalprobmirror1D}. However, the growth in the image term as $d$ increases causes $P(\Omega)$ to eventually overtake the free-space result instead of asymptotic to it.
The comparison is shown in Figure~\ref{fig: scaling1d}.  We also note that this growth is strictly a finite-time effect, i.e. not in the long interaction regime\footnote{We thank the anonymous referee for pointing this out.}. To see this, note that the total probability in Eq.~\eqref{eq:totalprobmirror1D} can be re-expressed as \textit{total probability} \textit{per unit time} $P(\Omega)/\sigma$ and in the limit $\sigma\to\infty$ we get 
\begin{align}
    \lim_{\sigma\to\infty}\frac{P(\Omega)}{\sigma} = - \frac{\sqrt{\pi}\Theta(-\Omega)}{\Omega} e^{2\Omega\epsilon}\rr{1- \cos (2\Omega d)}\,,
\end{align}
which is in fact oscillatory for excited detector ($\Omega<0$), as one may expect from e.g. \cite{Hodgkinson:2013tsa}. However, the transition probability itself diverges (though the \textit{rate} does not), so in this sense the large-$d$ growth of $P(\Omega)$ is similar to large-$\sigma$ growth, as one may have guessed from relativistic consideration which puts space and time on equal footing.

    \begin{figure}[tp]
        \centering
        \includegraphics[scale=1]{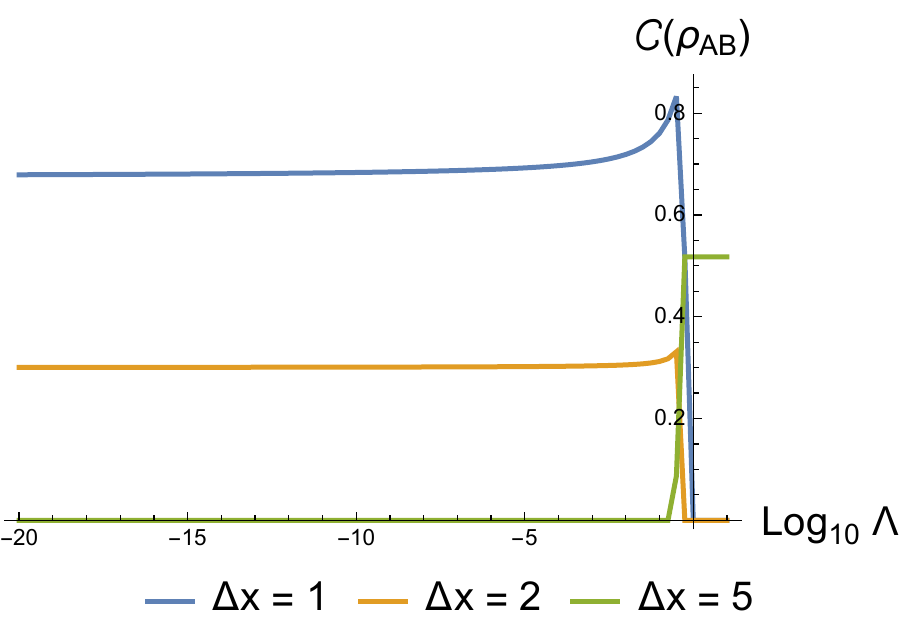}
        \caption{The concurrence in free space as a function of IR cutoff (in logarithmic scale), for $\Omega=1$ and $ \sigma=1$ in natural units (i.e. $\Omega\sigma=1$ is dimensionless).}
        \label{fig: IRcutoff}
    \end{figure}

    What this highlights is that within perturbation theory, one should take extra care in distinguishing what is due to the physics of the boundary and what is an artifact of the dimensionality of spacetime. An effect strictly confined to $(1+1)$ dimensions will not necessarily provide guidance to our understanding of $(3+1)$ dimensional physics. In particular the observable $P(\Omega)$ in the presence of a mirror is generically an ill-behaved quantity in (1+1) dimensions, even without the IR ambiguity associated with the free-space counterpart.
    
    However, not all is lost. As we saw earlier, the concurrence is a quantity based on the \textit{difference} $|X|-\sqrt{P_AP_B}$. Roughly speaking, we can understand this by noting that for static mirror we have a $\log[(\Delta\tau-i\epsilon)^2 - 4d_A^2]$ dependence due to the form of the Wightman $W(\sx_A,\sx_A')$. Very roughly, for $|X|$ we will have a contribution of the form 
    \begin{equation}
    \begin{split}
        &W(\sx_A,\sx_B')\\
        &\hspace{0.5cm}\sim\log\left[\rr{(\Delta\tau-i\epsilon) - (2d_A-\Delta x_{AB})^2}\right]
    \end{split}
    \end{equation}
    which can be seen to scale the same way as $\sqrt{P_AP_B}$ for any fixed $x_{AB}$.
    From this, we expect that concurrence $\mathcal{C}$ is in some sense `dimension-independent': it removes the part of the Wightman function that depends on the \textit{absolute distance} $d_A,d_B$ away from the mirror. What remains is the \textit{relative distance} dependence $\Delta x_{AB}$ of $\concur$ which leads to the usual decay of correlations as we make the detectors have larger spacelike separation.
    Effectively, $\concur$ `regulates' the ill-behaved part of $|X|$ and $P(\Omega)$ somewhat analogous to adding counterterms in QFT. This provides strong evidence that we can rely on concurrence calculations in $(1+1)$ dimensions even if we cannot separately trust the entries in the evolved density matrix $\rho_{AB}$ as $d\to\infty$ due to the divergent behavior at large mirror distance.
    
    With some thought, this growth of $P$ with distance from the mirror $d$ may not be as surprising at it seems: it is already known in classical gravity that the Newtonian gravitational potential in one-dimensional space also grows linearly with separation between masses \cite{Mann:1992ar,Ohta:1996wq} 
    unlike the $1/r$ power law in the three-dimensional context \cite{Romero1994}. We should also remark that since this is an inherent problem associated with the long-distance behaviour of the Wightman function, the issue will not disappear for general choices of ray-tracing functions. We can check this numerically for the three ray tracing functions in this paper -- we find that indeed $\concur$ seems to provide proper regularization in $(1+1)$ dimensions: that is, for each ray tracing function we find similar divergences in both the nonlocal term $|X|$ and probability $P$, but not in $\concur$. 
    
     As a consequence of this ``regularization'',  the concurrence $\concur$ also does not suffer from an IR ambiguity even in free space. For the same reason as why the divergences in the $d_A\to\infty$ limit is subtracted off by the definition of $\concur$, the IR ambiguity is also subtracted off. We can thus think of $\concur$ as having an IR regulator \textit{by definition}. This is the reason why the free space limit of the concurrence in the presence of mirrors is well-defined (cf. Figure~\ref{fig: distfromStatic}). Since concurrence is IR-regular, it can be reverse-engineered to induce a natural IR cutoff for the free-space Wightman function. This is shown in Figure~\ref{fig: IRcutoff}. We see that the concurrence approaches an approximately constant value once $\Lambda$ is reasonably small, which we can take to be at least $\Lambda\lesssim 10^{-3}$. In \cite{Pozas2015} and \cite{Eduardo2015firewall} the value $\Lambda=10^{-3}$ was adopted arbitrarily due to the ambiguity argument: this value is arguably small enough for their purposes, though choosing a smaller value allowed by the numerical precision used during computation is  preferable. We verified this for up to $\Lambda\sim 10^{-100}$ and the IR ambiguity does not seem to play a role in computation of $\concur$.

    In short, we can think of concurrence $\concur$ as a naturally ``regularized'' quantity for a massless scalar field in all dimensions, unlike the IR-divergent $|X|,P$ in $(1+1)$ dimensions with linear coupling. Furthermore, the regularity of $\concur$ may be instead used to induce a natural IR cutoff $\Lambda$ for the detector joint density matrix $\rho_{AB}$.

\bibliographystyle{JHEP}

\bibliography{myref}
\end{document}